\documentclass[12pt]{iopart}
\usepackage{iopams}
\usepackage{epsfig}
\input{epsf}

\begin{document}
\title[Cosmological Landscape From Nothing]{ Cosmological Landscape From Nothing: Some Like It Hot}
\author{A O Barvinsky$^{1}$ and A Yu Kamenshchik$^{2,3}$}
\address{$^{1}$\em Theory Department, Lebedev
Physics Institute, Leninsky Prospect 53, 119991 Moscow, Russia\\
$^{2}$Dipartimento di Fisica and INFN, Via Irnerio 46, 40126
Bologna, Italy\\
$^{3}$L.D.Landau Institute for Theoretical Physics, Kosygin Str. 2,
119334 Moscow, Russia}
\eads{\mailto{barvin@lpi.ru}, \mailto{kamenshchik@bo.infn.it}}
\begin{abstract}
We suggest a novel picture of the quantum Universe --- its creation
is described by the {\em density matrix} defined by the Euclidean
path integral. This yields an ensemble of universes --- a
cosmological landscape --- in a mixed state which is shown to be
dynamically more preferable than the pure quantum state of the
Hartle-Hawking type. The latter is dynamically suppressed by the
infinitely large positive action of its instanton, generated by the
conformal anomaly of quantum fields within the cosmological
bootstrap (the self-consistent back reaction of hot matter). This
bootstrap suggests a solution to the problem of boundedness of the
on-shell cosmological action and eliminates the infrared catastrophe
of small cosmological constant in Euclidean quantum gravity. The
cosmological landscape turns out to be limited to a bounded range of
the cosmological constant $\Lambda_{\rm min}\leq \Lambda \leq
\Lambda_{\rm max}$. The domain $\Lambda<\Lambda_{\rm min}$ is ruled
out by the back reaction effect which we analyze by solving
effective Euclidean equations of motion. The upper cutoff is
enforced by the quantum effects of vacuum energy and the conformal
anomaly mediated by a special ghost-avoidance renormalization of the
effective action. They establish a new quantum scale $\Lambda_{\rm
max}$ which is determined by the coefficient of the topological
Gauss-Bonnet term in the conformal anomaly. This scale is realized
as the upper bound --- the limiting point of an infinite sequence of
garland-type instantons which constitute the full cosmological
landscape. The dependence of the cosmological constant range on
particle phenomenology suggests a possible dynamical selection
mechanism for the landscape of string vacua.
\end{abstract}

\pacs{04.60.Gw, 04.62.+v, 98.80.Bp, 98.80.Qc}
\submitto{JCAP}
\maketitle
\section{Introduction}
The ideas of quantum cosmology \cite{Bryce,VilNB} and Euclidean
quantum gravity \cite{HH,H} are again attracting attention. One of
the reasons is the fact that the landscape of string vacua is too
big \cite{landscape} to hope that a reasonable selection mechanism
for the plethora of these vacua can be successfully worked out
within string theory itself. Thus, it is expected that other methods
have to be invoked, at least some of them appealing to the
construction of the cosmological wavefunction
\cite{OoguriVafaVerlinde,Tye1,Tye23,Brustein}.

This approach is based on the idea of quantum tunneling from the
classically forbidden state of the gravitational field.
Semiclassically this state is described in terms of the imaginary
time, that is by means of the Euclidean spacetime, so that the
corresponding amplitudes and probabilities are weighted by the
exponentiated Euclidean gravitational action, $\exp(-S_{\rm E})$.
The action is calculated on the gravitational instanton -- the
saddle point of an underlying path integral over Euclidean
4-geometries. This instanton gives rise to the Lorentzian signature
spacetime by analytic continuation across minimal hypersurfaces, and
the instanton topology allows one to interpret this continuation
either as a quantum tunneling or creation of the Universe from
``nothing". Then the minima of $S_{\rm E}$ give most probable values
of the physically interesting parameters of the Universe, including
initial conditions and fundamental coupling constants
\cite{Hawking0,big-fix}. Apparently, when applied to the string
landscape this framework can serve as a selection mechanism for the
most probable physical vacua.

There is an immediate difficulty with this program, which is rooted
in the problem of unboundedness of the Einstein action. The most
important example of the creation from ``nothing" is the
Hartle-Hawking wave function of the Universe \cite{HH,H}, which
describes nucleation of the de Sitter Universe from the Euclidean
4-dimensional hemisphere,
    \begin{eqnarray}
    \Psi_{\rm HH}\sim \exp(-S_{\rm E})
    =\exp(3\pi /2G\Lambda).                         \label{1}
    \end{eqnarray}
It has a negative action which diverges to $-\infty$ for the
cosmological constant $\Lambda\to 0$. This means that the vanishing
cosmological constant is infinitely more probable than a positive
one -- the result which has a very controversial status. On one hand
it was used as a justification of a zero value of $\Lambda$
\cite{Hawking0} by the Coleman big-fix mechanism \cite{big-fix}, but
on the other hand it was considered as extremely confusing and
anti-intuitive because it preferred creation of infinitely large
universes and disfavored inflation. No need to say that this result
also looks unappropriate in context of the cosmological acceleration
which requires very small, but still nonvanishing cosmological
constant (certainly, if we identify dark energy phenomenon with the
cosmological constant effect).

Such a controversy gave rise to the so-called cosmology debate
\cite{debate} reflecting the first attempt to circumvent this
problem. Point is that the alternative formalism of the
Wheeler-DeWitt equation, which is only formally equivalent to the
path integral approach \cite{Leutwyler,BarvU}, admits also an
opposite sign in the exponential of (\ref{1}). It exists either in
the form of Linde \cite{Linde} or Vilenkin \cite{Vilenkin} tunneling
proposals. These proposals differ by boundary conditions
\cite{debate} but qualitatively render the same exponentially small
probability for $\Lambda\to 0$, which opens the possibility for
conclusions opposite to the Hartle-Hawking case. In particular, the
inclusion of one-loop effects allows one to shift most probable
values of the effective cosmological constant from zero to a narrow
highly peaked range compatible with the requirements of inflation
\cite{scale}.

Such a big discrepancy in handling this infrared domain gives a
feeling of indeterminacy -- fundamentals of quantum gravitational
tunneling seem to be on too shaky a ground to serve as a reliable
completion of the string theory of everything. Apparently, this
discrepancy could be resolved within a deeper understanding of the
gravitational path integral and correct operator realization of the
Wheeler-DeWitt equation. However, this achievement might still be in
vain, because in view of the anticipated string nature of everything
this equation might be not fundamental. For this and other reasons
(we will not repeat here usual arguments refuting the tunneling
prescription on the basis of erroneousness of anti-Wick rotation,
etc.) we will stick to the Hartle-Hawking prescription of the
Euclidean path integration.

In this prescription the attempts to resolve the infrared
catastrophe of small $\Lambda$ are very hard to implement. Indeed,
within a conventional wisdom of a low-derivative expansion all
quantum corrections are of higher order in the curvature $R$ than
the cosmological and Einstein terms. Therefore, they are negligibly
small in the infrared limit $R\sim\Lambda\to 0$ and cannot compete
with the infinitely growing tree-level contribution of (\ref{1}).
This is a very general argument that cannot be circumvented within
the local curvature expansion of the effective action. So the only
reasonable option is to try essentially {\em nonlocal} quantum
effects mediated by nonlocal terms of non-vacuum nature.

This idea was recently attempted in speculative but very
thought-provoking papers of Henry Tye {\em et al} \cite{Tye1,Tye23}
who suggested to consider the effect of radiation in quantum
creation from nothing (see also \cite{Brustein}). Radiation
effectively appeared in \cite{Tye1,Tye23} as a result of ultraviolet
(and infrared) renormalization of the effective action of the
``environment" -- spatially inhomogeneous modes of matter and
gravitational fields, treated perturbatively on the minisuperspace
FRW background. Their contribution to the instanton action was
advocated in \cite{Tye1,Tye23} to be positive and inverse
proportional to the square of $\Lambda$, so that the action was
bounded from below at some positive $\Lambda$ with which the
Universe gets created from ``nothing". This mechanism has already
been justifiably criticized in \cite{Mersini} by noting that this
type of ``radiation" can only renormalize the effective value of
$\Lambda$, in terms of which the boundedness properties of the
action remain the same. Nevertheless, as we show below, the
radiation contents of the creation-from-nothing instanton can really
mediate the boundedness of its effective action. However, the origin
of this radiation and the mechanism of the infrared cutoff at small
$\Lambda$ are drastically different from those of \cite{Tye1,Tye23}.
\\
\begin{figure}[h]
\centerline{\epsfxsize 7.5cm \epsfbox{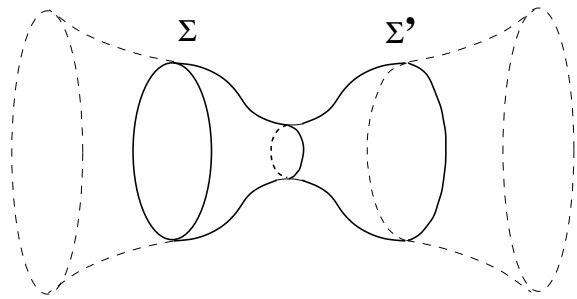}} \caption{\small
Picture of instanton representing the density matrix. Dashed lines
depict the Lorentzian Universe nucleating from the instanton at the
minimal surfaces $\Sigma$ and $\Sigma'$. \label{Fig.1}}
\end{figure}

The core of our suggestion is a simple observation that the presence
of radiation implies a statistical ensemble described by the density
matrix, rather than a pure state assumed in \cite{Tye1,Tye23}.
Density matrix in Euclidean quantum gravity \cite{Page},
$\rho[\,\varphi,\varphi']$, originates from an instanton with two
disjoint boundaries $\Sigma$ and $\Sigma'$ associated  respectively
with its two entries $\varphi$ and $\varphi'$ (collecting both
gravitational and matter variables). The bulk part of the instanton
interpolates between these boundaries and, thus, establishes mixing
correlations between the observables located at them, see
Fig.\ref{Fig.1}.
\begin{figure}[h]
\centerline{\epsfxsize 8cm \epsfbox{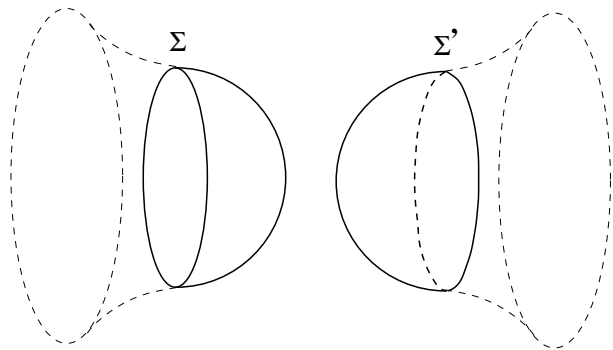}} \caption{\small
Density matrix of a pure state represented by the union of two
vacuum Hartle-Hawking instantons giving rise to Lorentzian
spacetime. \label{Fig.2}}
\end{figure}
In contrast, the pure density matrix of the Hartle-Hawking state
corresponds to the situation when the instanton bridge between
$\Sigma$ and $\Sigma'$ is broken, so that topologically the
instanton is a union of two disjoint hemispheres. Each of the
half-instantons smoothly closes up at its pole which is a regular
internal point of the Euclidean spacetime ball (Fig.\ref{Fig.2})
--- a picture illustrating the factorization of
$\hat\rho=|\Psi_{\rm HH}\rangle\langle\Psi_{\rm HH}|$.

The concept of an initial state of the Universe should be extended
to include mixed states of the above type. This extension seems
natural because it is up to Nature to decide whether a pure or mixed
initial state is preferable. So this is a dynamical question rather
than a postulate, and we address this question below by
incorporating both types of states into a unified framework of the
density matrix.\footnote{It should be emphasized that, unlike in
black hole theory and the deSitter cosmology with horizons, mixed
nature of the quantum state is fundamental here, because it is not
effectively acquired as a result of tracing out invisible
superhorizon degrees of freedom or coarse graining. This is just a
fundamental lack of physical information about the quantum state
which should be described in terms of a density matrix rather than a
pure quantum state. This situation is not postulated but treated on
equal footing with the case of a pure quantum state carrying the
full amount of physics information admissible by the principles of
quantum mechanics. As we will see below, the system dynamically
prefers mixed states of quasi-thermal nature.} This is attained by
considering the density matrix as a Euclidean path integral over the
fields on spacetimes with a tubular topology of Fig.\ref{Fig.1},
which includes as a limiting case the topology of Fig.\ref{Fig.2}
with a broken bridge.

\begin{figure}[h]
\centerline{\epsfxsize 12cm \epsfbox{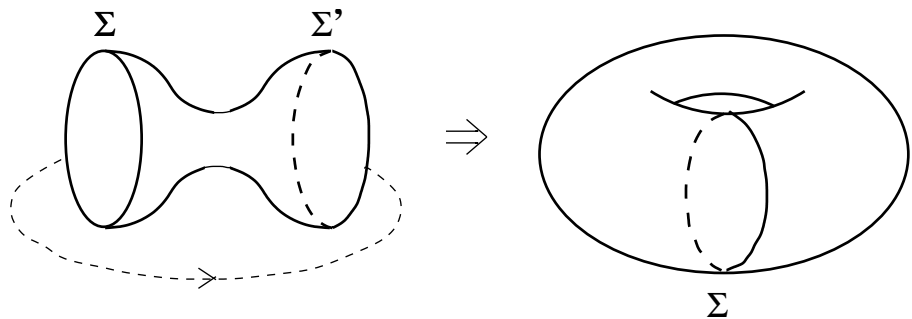}}\caption{\small
Calculation of the partition function represented by
compactification of the instanton to a torus with periodically
identified Euclidean time.\label{Fig.3}}
\end{figure}

Qualitatively, the main effect that we advocate here is as follows.
When calculated in the saddle-point approximation the density matrix
automatically gives rise to radiation whose thermal fluctuations
destroy the Hartle-Hawking instanton. Namely, the radiation stress
tensor prevents the half-instantons of the above type from closing
and, thus, forces the tubular structure on the full instanton
supporting the thermodynamical nature of the physical state. The
existence of radiation, in its turn, naturally follows from the
partition function of this state. The partition function originates
from integrating out the field $\varphi$ in the coincidence limit
$\varphi'=\varphi$. This corresponds to the identification of
$\Sigma'$ and $\Sigma$, so that the underlying instanton acquires
toroidal topology, see Fig.\ref{Fig.3}. Its points are labeled by
the periodically identified Euclidean time, a period being related
to the inverse temperature of the quasi-equilibrium radiation. The
back reaction of this radiation supports the instanton geometry in
which this radiation exists, and we derive the equation which makes
this bootstrap consistent.
\\

We show that for the radiation of conformally-invariant fields the
analytical and numerical solution of the bootstrap equations yields
a set of instantons --- a landscape --- only in the bounded range of
$\Lambda$,
    \begin{eqnarray}
    \Lambda_{\rm min}<\Lambda<\Lambda_{\rm max}.            \label{2}
    \end{eqnarray}
This set consists of the countable sequence of one-parameter
families of instantons which we call garlands, labeled by the number
$k=1,2,3,...$ of their elementary links (of the form shown on
Fig.\ref{Fig.1}). Each of these families spans a continuous subset,
$\Lambda_{\rm min}^{(k)}<\Lambda\leq\Lambda_{\rm max}^{(k)}$,
belonging to (\ref{2}). These subsets of monotonically decreasing
lengths $\Lambda_{\rm max}^{(k)}-\Lambda_{\rm min}^{(k)}\sim 1/k^4$
do not overlap and their sequence has an accumulation point at the
upper boundary $\Lambda_{\rm max}$ of the range (\ref{2}). Each of
the instanton families at its upper boundary $\Lambda_{\rm
max}^{(k)}$ saturates with the static Einstein Universe filled by a
hot equilibrium radiation with the temperature $T_{(k)}\sim m_P/\ln
k^2$, $k\gg1$, and having the {\em negative} decreasing with $k$
action $\Gamma_0^{(k)}\sim -\ln^3k^2/k^2$. This means that large
$k$ garlands do not dominate the cosmological landscape and turn out
to be less probable than $k=1$ instantons.

Remarkably, all values of $\Lambda$ below the range (\ref{2}),
$\Lambda<\Lambda_{\rm min}$, are completely ruled out either because
of the absence of instanton solutions or because of their {\em
infinitely large positive} action. This effect is provided by the
radiation back reaction and survives even in the classical limit of
the theory. A similar situation holds for $\Lambda>\Lambda_{\rm
max}$ -- no instantons exist there, and the Lorentzian
configurations in this overbarrier domain (if any) are exponentially
suppressed relative to those of (\ref{2}). However, the high-energy
cutoff is entirely the quantum effect of the vacuum energy of
quantized fields and their conformal anomaly. This effect generates
a new scale $\Lambda_{\rm max}$ in gravity theory. Specifically,
this scale is determined by the coefficient of the (topological)
Gauss-Bonnet term in the conformal anomaly.

Thus, neither infrared nor ultraviolet catastrophe is present in our
picture. Rather, we get a bounded cosmological landscape. Its
boundaries depend on the phenomenology of quantum fields. In
particular, for a single conformal scalar and vector field they are
respectively given by the following Planckian values, $m_P^2\equiv
3\pi/4G$,
    \begin{eqnarray}
    &&\Lambda_{\rm min}^{\rm scalar}
    \approx 8.991\,m_P^2,\,\,\,
    \Lambda_{\rm max}^{\rm scalar}=360\,m_P^2, \label{3}\\
    &&\Lambda_{\rm min}^{\rm vector}
    \approx 3.189\,m_P^2,\,\,\,
    \Lambda_{\rm max}
    ^{\rm vector}=180/31\,m_P^2.  \label{4}
    \end{eqnarray}
Qualitatively, these bounds decrease with the growth of the field
spin and also go to zero as $1/N$ with the growing number of fields
$N$. This brings us to the domain of semilassical consistency of the
model. This also means that the landscape shrinks to a narrower and
narrower subplanckian range when we ascend the phenomenology energy
scale. This suggests the possibility of generating a long sought
selection mechanism which might disentangle from the vast string
landscape a dynamically preferred vacuum compatible with the
observed Universe.

The paper is organized as follows. In Sect.2 we present the effect
of radiation in Euclidean cosmology which maintains a thermal nature
of the cosmological instanton and underlies the density-matrix
entanglement. In Sect.3 we construct the path integral
representation for this density matrix and the corresponding
partition function (or the Euclidean effective action). Here we also
describe the approximation when the minisuperspace metric of the
instanton is subject to the self-consistent back reaction of quantum
matter. In Sects.4-6 we calculate the effective action of
conformally invariant fields on the instanton with generic FRW
metric and derive the nonlocal effective equation for the latter.
The effective action is obtained as a sum of contributions of the
conformal anomaly of quantum fields and their action on the static
Einstein Universe. They are both properly renormalized to avoid
ghost instabilities in the solutions of the corresponding effective
equation. The latter turns out to be the Friedmann equation which is
modified by the conformal anomaly term and has as a source the
radiation density term determined by the bootstrap equation.
Elimination of the infrared catastrophe by the back-reaction effect
of a conformal matter is demonstrated within this bootstrap in
Sect.7. The infinite sequence of garlands -- multiply folded
instantons generating the new quantum scale as an upper bound of the
landscape (\ref{2}) -- is constructed in Sect.8. The concluding
section summarizes the results and discusses the extension of the
model to quantum fields of higher spins and a possible application
of the bounded cosmological landscape as a means of reducing the
landscape of string vacua.

\section{The effect of radiation}
The effect of radiation on cosmological solutions of Euclidean
gravity is as follows. In a spatially closed cosmology with the
Euclidean FRW metric ($\Omega^{(3)}$ is a 3-sphere of a unit radius)
    \begin{equation}
    ds^2 = N^2(\tau)\,d\tau^2 +
    a^2(\tau)\,d^2\Omega^{(3)}, \label{metric}
    \end{equation}
the Friedmann equation includes the radiation density term inverse
proportional to the fourth power of the scale factor $a$
    \begin{equation}
    \frac{\dot{a}^2}{a^2} =
    \frac{1}{a^2} - H^2 -\frac{C}{a^4}.     \label{Friedmann0}
    \end{equation}
Here we choose the gauge of a unit lapse $N=1$, $H^2=\Lambda/3$ is
the cosmological constant in terms of the Hubble parameter $H$ and
the constant $C$ characterizes the amount of radiation. The
radiation term blows up to infinity for $a\to 0$ and in view of the
positivity of the radiation density ($C>0$) prevents the model from
the cosmological collapse. Moreover, the right hand side here also
vanishes for some big $a$, and the Friedmann equation has two
turning points -- the situation analyzed in the context of quantum
tunneling in \cite{Rubakov,Khalatnikov}. In the corresponding
solution \cite{Halliwell}
    \begin{equation}
    a(\tau) = \frac{1}{\sqrt{2}H}
    \sqrt{1-(1-4CH^2)^{1/2}\cos
    2H\tau}                                     \label{instanton}
    \end{equation}
the scale factor never goes to zero and varies between the maximal
and minimal values
    \begin{equation}
    a_\pm= \frac{1}{\sqrt{2}H}
    \sqrt{1\pm(1-4CH^2)^{1/2}}     \label{apm}
    \end{equation}
during one period of the Euclidean time $-\pi/2H\leq\tau\leq\pi/2H$.
This instanton solution exists for a finite amount of radiation
satisfying the bound
    \begin{equation}
    4H^2C \leq 1,                              \label{inequality}
    \end{equation}
and it can be used for constructing the density matrix of the
Universe if we identify the boundaries $\Sigma$ and $\Sigma'$ with
the minimal surfaces of the maximal size $a_+$ at $\tau=\pm\pi/2H$.
The Lorentzian spacetime with the quasi-deSitter metric nucleates
from them via the analytic continuation in time and gives rise to
inflationary cosmology with the asymptotic value of the Hubble
constant $H$ at late stages of expansion, see Fig.\ref{Fig.1} above.
As we see, at any nonzero $H$ the radiation term prevents the
instanton bridge between $\Sigma$ and $\Sigma'$ from breaking and
thus supports the thermal nature of the ensemble of universes.

Note that the (half-period) instanton interpolating between the
minimal and maximal values of a scale factor, which was used in
\cite{Rubakov, Khalatnikov} for the description of quantum tunneling
between different Lorentzian domains, is not suitable for the
density matrix. The calculation of thermodynamic averages
incorporates taking the trace and leads to the identification of
$\Sigma$ and $\Sigma'$ surfaces. Therefore, these surfaces should
have the same geometric characteristics. Thus, only full-period and
multiple-period instantons are admissible.

\section{The density matrix}
The picture of the above type serves as a tree-level approximation
for the density matrix of this ensemble. At the exact level this
density matrix is given by the Euclidean path integral
    \begin{eqnarray}
    \rho[\,\varphi,\varphi'\,]=\mbox{\large$e$}^{\,\textstyle
    \Gamma}\!\!\!\!\!\!\!\!\!\!\!\!\!\!
    \int\limits_{\,\,\,\,\,\,\,\,g,\,
    \phi\,|_{\,\Sigma,\Sigma'}\,=\,(\,\varphi,\varphi')}
    \!\!\!\!\!\!\!\!\!D[\,g,\phi\,]\,
    \exp\big(-S_{\rm E}[\,g,\phi\,]\big).             \label{rho}
    \end{eqnarray}
Here $S_{\rm E}[\,g,\phi\,]$ is the classical action of the theory,
and the integration runs over gravitational $g$ and matter $\phi$
fields on a tubular spacetime interpolating between the field
configurations $\varphi$ and $\varphi'$ at $\Sigma$ and $\Sigma'$
(which, as mentioned above, also include both gravitational and
matter variables). The normalization factor $\exp(\Gamma)$ is
determined from the condition ${\rm tr}\,\hat\rho=1$. This condition
serves as a definition of the Euclidean effective action
    \begin{eqnarray}
    \mbox{\large$e$}^{\textstyle -\Gamma}=\!\!\!\!\!
    \int\limits_{\,\,\,\,\,
    g,\,\phi\,|_{\Sigma{\vphantom'}}\,
    =\,g,\,\phi\,|_{\Sigma'}}
    \!\!\!\!\!\!D[\,g,\phi\,]\,
    \exp\big(-S_{\rm E}[\,g,\phi\,]\big)           \label{pathGamma}
    \end{eqnarray}
on the compactified spacetime obtained by identifying the boundaries
$\Sigma$ and $\Sigma'$ as shown on Fig.\ref{Fig.3} above. Now the
integration runs over periodically identified gravity and matter
fields.

The motivation for this definition of the density matrix and its
statistical sum can be found in \cite{Page}. Here we would only
mention that this is a natural generalization of the path integral
for the no-boundary wave function of the Universe. Also, it can be
regarded as a gravitational generalization of the density matrix of
the equilibrium thermodynamical ensemble at finite temperature
$T=1/\beta$, $\hat\rho=\exp\big(\Gamma-\beta\hat H\big)$, for a
system with a Hamiltonian operator $\hat H$. Its kernel in the
functional coordinate representation
$\rho[\varphi,\varphi']=\langle\varphi\,|\,\hat\rho\,|\,\varphi'\rangle$,
similarly to (\ref{rho}), is given by the Euclidean path integral
over histories $\phi(\tau)$ in the imaginary time $\tau$,
interpolating between $\phi(0)=\varphi'$ and $\phi(\beta)=\varphi$.

This definition suggests an unusual interpretation of the
cosmological ensemble. It follows that, semiclassically, the set of
Lorentzian spacetime universes originates by analytic continuation
into the complex plane of the temperature variable rather than the
time variable. Therefore, the whole picture looks more like a
thermal state of an already existing universe, rather than its
quantum birth from ``nothing" by means of tunneling from the
classically forbidden state. In fact, this situation is a
manifestation of the principle that ``the mathematical formalism 
 [of the quantum theory] 
is
capable of yielding its own interpretation" put forward by DeWitt in the
other context \cite{DW1}. Indeed, one and the same formalism of the
imaginary time is used to describe qualitatively different phenomena
--- thermodynamics and quantum tunneling. On the physical grounds
they are easily distinguishable in usual field-theoretic problems,
but here they exhibit intrinsic dualism between thermodynamics and
quantum tunneling.\footnote{Similarly, the formalism of quantum
black holes can be interpreted in terms of widely accepted black
hole thermodynamics and much less recognized physics of virtual
black holes \cite{virtualBH}.}

Anyway, the density matrix (\ref{rho}) unambiguously prescribes a
particular mixed quantum state of the system. Its semiclassical
calculation yields as a saddle point the configuration of a tubular
topology depicted on Fig.\ref{Fig.1}. Its Euclidean part
(interpreted either in thermodynamical or tunneling terms) is
bounded by two minimal surfaces $\Sigma$ and $\Sigma'$. The analytic
continuation across these surfaces yields the ensemble of
cosmological models expanding in real Lorentzian time, and this
picture can be called as the origin of cosmological thermodynamics
via creation from ``nothing". Taking the trace of $\hat\rho$ in the
normalization condition for the density matrix results in the
identification of $\Sigma$ and $\Sigma'$ and the toroidal
compactification of the instanton depicted on Fig.\ref{Fig.3}. This
underlies the semiclassical calculation of the statistical sum and
the corresponding Euclidean effective action (\ref{pathGamma}).

The thermal quasi-equilibrium nature of this statistical sum will be
demonstrated by explicit calculations in Sect.5 below. However, the
thermodynamical analogy for the density matrix kernel, that was
mentioned above, is not entirely precise here due to general
covariance of gravity theory. This happens because: i) the
Hamiltonian of a spatially closed cosmology is vanishing and ii) the
temperature in (\ref{rho}) is not specified by hands as a free
parameter, but rather integrated over (as an integration over the
lapse of Euclidean time separating the surfaces $\Sigma$ and
$\Sigma'$). Within semiclassical expansion the second property works
as follows. The effective ``temperature" of the instanton (or its
Euclidean time period) is uniquely determined from its equation of
motion -- Euclidean Friedmann equation for a cosmological scale
factor.

This equation and the way it incorporates back reaction of quantum
matter arise as follows. We decompose the full configuration space
of $[\,g,\phi\,]$ in (\ref{pathGamma}),
$[\,g,\phi\,]\to[\,g_0(\tau),\Phi(x)\,]$, into a minisuperspace
of the cosmological scale factor $a(\tau)$ and the lapse function
$N(\tau)$, $g_0(\tau)=\big(a(\tau),N(\tau)\big)$, and the sector of
"matter" $\Phi(x)$ which includes together with matter fields
also the metric perturbations on the minisuperspace background
(\ref{metric}), $\Phi(x)=(\phi(x),\psi(x),A_\mu(x),
h_{\mu\nu}(x),...)$. Under this decomposition the measure factorizes
accordingly $D[\,g,\phi\,]=Dg_0(\tau) \times D\Phi(x)$ (both
factors implying relevant gauge-fixing procedures in minisuperspace
and field sectors respectively), and the statistical sum
(\ref{pathGamma}) takes the form
    \begin{eqnarray}
    \mbox{\large$e$}^{\textstyle
    -\Gamma}=
    \int Dg_0(\tau)\,
    \exp\Big(\!-\Gamma[\,g_0(\tau)\,]\Big),        \label{intg0}
    \end{eqnarray}
where the effective action $\Gamma[\,g_0(\tau)\,]$ of quantized
matter on the minisuperspace background $g_0(\tau)$ is determined by
the path integral
    \begin{eqnarray}
    &&\mbox{\large$e$}^{\textstyle
    -\Gamma[\,g_0(\tau)\,]}=
    \int D\Phi(x)\,
    \exp\Big(\!-S_{\rm E}[\,g_0(\tau),\Phi(x)\,]\Big).
    \end{eqnarray}

Our approximation in what follows will be to include into this
action the one-loop order,
    \begin{eqnarray}
    &&\Gamma[\,g\,]=S_{\rm E}[\,g\,]
    +\Gamma_{\rm 1-loop}[\,g\,],
    \end{eqnarray}
and calculate the minisuperspace integral (\ref{intg0}) in the
tree-level approximation, which reduces to solving the {\em
effective equations} for $\Gamma[\,g_0(\tau)\,]$. This will give
the lowest order back reaction effect of quantum matter in the
process of quantum creation of the cosmological ensemble.

\section{Conformal anomaly and ghosts}
Even for a simple FRW background the one-loop action is not exactly
calculable for quantum fields of a general type. For
conformally-invariant fields, however, one can apply the technique
of the conformal transformation which relates the FRW metric
(\ref{metric}) rewritten in terms of the conformal time $\eta$,
    \begin{equation}
    ds^2 = a^2(\eta)(d\eta^2 + d^2\Omega^{(3)}), \label{metric3}
    \end{equation}
to the metric of the Einstein static Universe of a unit size
    \begin{equation}
    d\bar s^2 = d\eta^2 + d^2\Omega^{(3)}. \label{metric4}
    \end{equation}

In contrast to the local conformal invariance of the classical
action of a conformal field of weight $w$,
    $S^{\rm conf}[\,e^\sigma g_{\mu\nu}(x),e^{-w\sigma/2}\phi(x)\,]=
    S^{\rm conf}[\,g_{\mu\nu}(x),\phi(x)\,]$, $\sigma=\sigma(x)$,
its quantum effective action has a conformal anomaly
    \begin{equation}
    g_{\mu\nu}\frac{\delta
    \Gamma_{\rm 1-loop}}{\delta g_{\mu\nu}} =
    \frac{1}{4(4\pi)^2}g^{1/2}
    \left(\alpha \Box R +
    \beta E + \gamma C_{\mu\nu\alpha\beta}^2\right),    \label{anomaly}
    \end{equation}
determined by the coefficients of the scalar $\Box R$, Weyl tensor
squared term $C^2_{\mu\nu\alpha\beta}$ and the Gauss-Bonnet
invariant
    \begin{eqnarray}
    &&E = R_{\mu\nu\alpha\gamma}^2
    -4R_{\mu\nu}^2 + R^2.                     \label{Euler}
    \end{eqnarray}
These coefficients depend on the spin of a conformal field
\cite{confanomaly} and read respectively for a scalar, Weyl spinor
and vector field as
    \begin{eqnarray}
    \alpha=\frac1{90}\times\left\{\begin{array}{c} -1 \\
    -3\\
    18\end{array}\right.\,\, ,
    \,\,\,\,\,
    \beta=\frac1{360}\times
    \left\{\begin{array}{cl} 2\\
    11\\
    124\end{array}\right.\,,                \label{100}
    \end{eqnarray}
where for the vector case we cite the result of the $\zeta$-function
\cite{DowkerCritchley} and point separation \cite{Christensen}
regularizations, which differs from the dimensional regularization
result \cite{BrownCassidy} for $\alpha$.

This anomaly, when integrated functionally along the orbit of the
conformal group, gives the relation between the actions on generic
conformally related backgrounds \cite{Tseytlinconf}
    \begin{eqnarray}
    &&\Gamma_{\rm 1-loop}[\,g\,]=
    \Gamma_{\rm 1-loop}[\,\bar g\,]+\Delta\Gamma[\,g,\bar
    g\,],\,\,\,g_{\mu\nu}(x)=e^{\sigma(x)}\bar g_{\mu\nu}(x),
    \end{eqnarray}
where \cite{BMZ}
    \begin{eqnarray}
    &&\Delta\Gamma[\,g,\bar g\,]=
    \frac{1}{2(4\pi)^2}\int d^4x \bar g^{1/2} \left\{\,\frac{1}{2}\,
    \Big[\,\gamma\, \bar C_{\mu\nu\alpha\beta}^2
    +\beta\,\Big(\bar E-\frac{2}{3}\,\bar\Box \bar R\Big)\Big]\,
    \sigma\right.                                  \nonumber\\
    &&\qquad\qquad\qquad
    \left.+\,\frac{\beta}{2}\,\Big[\,(\bar\Box\sigma)^2
    +\frac{2}{3}\,\bar R\,(\bar\nabla_{\mu}\sigma)^2\,\Big]\,\right\}\nonumber\\
    &&\qquad\qquad\qquad-
    \,\frac{1}{2(4\pi)^2}\Big(\frac{\alpha}{12}
    +\frac{\beta}{18}\Big)\,\int d^4x\,\Big(g^{1/2}R^2(g)-
    \bar{g}^{1/2}R^2(\bar{g})\Big),              \label{deltaW}
    \end{eqnarray}
and all barred quantities are calculated with respect to the metric
$\bar g_{\mu\nu}(x)$.

For the conformal factor $e^\sigma=a^2(\tau)$ this expression
immediately generates higher-order derivative terms $\sim \ddot a^2$
in the effective Lagrangian. This certainly produces ghost
instabilities in solutions of effective equations. It is remarkable,
however, that higher-derivative terms are all proportional to the
coefficient $\alpha$, because similar terms linear in $\beta$
completely cancel out in (\ref{deltaW}) ($(\bar\Box\sigma)^2$
against the $R^2(g)$-term). The $\alpha$-term of the conformal
anomaly can be arbitrarily changed by adding a {\em local}
counterterm $\sim g^{1/2}R^2$, admissible from the viewpoint of
general renormalization theory. Therefore we can put it identically
to zero, $\alpha\to\alpha_R=0$, by the following {\em
ghost-avoidance} renormalization of the effective action
    \begin{eqnarray}
    \Gamma_{\rm 1-loop}[\,g\,]\to \Gamma_{R}[\,g\,]
    =\Gamma_{\rm 1-loop}[\,g\,]
    +\frac1{2\,(4\pi)^2}\,
    \frac\alpha{12}\int d^4x\,
    g^{1/2}\,R^2(g).                   \label{renormalization}
    \end{eqnarray}
Certainly, this additionally spoils conformal invariance of the
theory which was anyway irreversibly broken by quantum corrections.
Thus it is reasonable to fix this {\em local} renormalization
ambiguity by the additional criterion of the absence of ghosts, what
we do here for sake of consistency of the theory at the quantum
level.\footnote{This is certainly not an exhaustive solution of the
ghost problem in effective equations of local field models.
Higher-derivative terms still remain in the other sectors of the
theory -- the graviton sector of transverse-traceless modes, in
particular. These sectors, however, are not involved within the
minisuperspace sector of a homogeneous FRW metric, and it is
suggestive to use a simple method of this finite ghost-avoidance
renormalization which ultimately eradicates ghosts in this sector.}

From (\ref{deltaW}) it then follows that the conformal contribution
to the {\em renormalized} action on the minisuperspace background
(\ref{metric3})-(\ref{metric4}) equals
    \begin{eqnarray}
    &&\Delta \Gamma[\,g,\bar g\,]\equiv
    \Gamma_{R}[\,g\,]
    -\Gamma_{R}[\,\bar g\,]=
    m_P^2\,B\!\int d\tau
    \left(\,\frac{\dot{a}^2}{a}
    -\frac16\,\frac{\dot{a}^4}a\right),   \label{correction}\\
    &&m_P^2\,B=\frac34\,\beta,       \label{BmP^2}
    \end{eqnarray}
with the constant $m_P^2\,B$ which for scalars, two-component
spinors and vectors equals respectively $1/240$, $11/480$ and
$31/120$. As we will see, the positivity of the first term in the
integrand of this expression will play a crucial role in the
dynamical suppression of the vacuum Hartle-Hawking type instantons.

\section{Effective action on a static Einstein instanton}
Now we need $\Gamma_{R}[\,\bar g\,]$ on a static background
(\ref{metric4}) with a periodically identified time $\eta$ having
the period $\eta_0$. This is a typical thermodynamical calculation
which, for completeness, we fully present here for a conformal
scalar field. By decomposing it in spherical harmonics (enumerated
by a collective index $n$) on a unit 3-sphere one gets the classical
action as a sum of (Euclidean) oscillators with energies $\omega_n$
    \begin{equation}
    S[\,\bar g,\phi\,] = \frac12\,
    \sum_{n}\int_0^{\eta_0}
    d\eta \left(\Big(\frac{d\phi_n}{d\eta}\Big)^2+
    \omega_n^2\,\phi^2_n\right).                   \label{scal-action}
    \end{equation}
The value of this action on solutions of classical equations
    \begin{eqnarray}
    &&\frac{d^2\phi_n}{d\eta^2}-\omega_n^2\phi_n = 0,
    \,\,\,\,n=1,2,...\,,                           \label{conf-eq}\\
    &&\phi_n(0)=\varphi'_n,\,\,\,
    \phi_n(\eta_0)=\varphi_n,
    \end{eqnarray}
gives the Hamilton-Jacobi function
    \begin{eqnarray}
    &&S(\varphi,\eta_0\,|\,\varphi',0) =\sum_n
    \omega_n\,\frac{(\varphi_n^2+\varphi'^2_n)\,
    \cosh (\omega_n\eta_0) -
    2\,\varphi_n\varphi'_n}
    {2\sinh(\omega_n\eta_0)}\,.         \label{scal-action1}
    \end{eqnarray}
Via the Pauli-Van Vleck formula it yields, up to a numerical factor,
the evolution operator in the Euclidean time $\eta_0$ -- the density
matrix
    \begin{eqnarray}
    &&\int\limits_{\,\,\,\,\,
    \phi_n(\eta_0)
    =\varphi_n,\; \phi_n(0)=\varphi'_n}
    \!\!\!\!\!\!\!\!D[\,\phi\,]\,
    \exp\big(-S[\,\bar g,\phi\,]\big)\nonumber\\
    &&\qquad\qquad\qquad\qquad=
    \left(\prod_n\frac{\partial^2
    S(\varphi,\eta_0\,|\,\varphi',0)}
    {\partial\varphi_n\;\partial\varphi'_n}\right)^{1/2}
    \exp\Big(-S(\varphi,\eta_0\,|\,\varphi',0)\Big).
    \end{eqnarray}
Its trace generates the needed effective action
    \begin{eqnarray}
    &&\mbox{\large$e$}^{\textstyle
    -\Gamma_{\rm 1-loop}[\,\bar g\,]}=\int \prod_n d\varphi_n
    \!\!\!\!\!\!\!\!\!\!\!\!\!\!\!\!\!\!\!\!\!\!
    \int\limits_{\,\,\,\,\,\,\,\,\,\,\,\,\,\,\,\,
    \,\,\,\,\,\,\,\,\,\,\,\,
    \phi_n(\eta_0)=\phi_n(0)=\varphi_n}
    \!\!\!\!\!\!\!\!\!\!\!\!\!\!\!\!\!\!
    \!\!\!\!\!\!
    D[\,\phi\,]\,
    \exp\big(-S[\,\bar
    g,\phi\,]\big)
    \nonumber\\
    &&\qquad\qquad\qquad\qquad\qquad\qquad\qquad\qquad\qquad
    ={\rm const}\,\prod_n\left(\sinh
    \frac{\omega_n\eta_0}2\right)^{-1},
    \end{eqnarray}
so that up to an additive constant it equals the sum of
contributions of the vacuum energy $E_0$ and free energy $F(\eta_0)$
    \begin{eqnarray}
    &&\Gamma_{\rm 1-loop}[\,\bar g\,]
    =\sum_{n}\left[\,\eta_0\,
    \frac{\omega_n}{2}
    +\ln\big(1-e^{-\omega_n\eta_0}\big)\,\right]=
    m_P^2\,E_0\,\eta_0+F(\eta_0),            \label{1000}\\
    &&m_P^2\,E_0=\sum_{n}
    \frac{\omega_n}{2}=\sum_{n=1}^\infty
    \frac{n^3}{2},                           \label{E_0}\\
    &&F(\eta_0)=\sum_{n}
    \ln\big(1-e^{-\omega_n\eta_0}\big)\equiv
    \sum_{n=1}^\infty n^2\,
    \ln\big(1-e^{-n\eta_0}\big).            \label{freeenergy}
    \end{eqnarray}
Here we go over to the explicit summation over the principal quantum
number $n=1,2,3,...$, in terms of which $\omega_n=n$ and $n^2$ is
the degeneracy of the energy level. Also we note that our definition
of the free energy $F(\eta_0)$ differs from the conventional one by
a factor of $\eta_0$.

Quartic divergence of the vacuum energy (\ref{E_0}) in
$\Gamma_{\rm 1-loop}[\,\bar g\,]$ constitutes the ultraviolet
divergences of the full action $\Gamma_{\rm 1-loop}[\,g\,]$.
Under a covariant regularization the power and quartic divergences
among them are absorbed by the renormalization of the cosmological
and Einstein terms, whereas the subtraction of logarithmic
divergences yields as a remnant the contribution of a conformal
anomaly considered above. For conformal fields, which we consider,
the logarithmic divergences are actually zero, because they are
given by the sum of integrated Weyl-squared and Euler number terms
($\gamma$ and $\beta$ terms of the conformal anomaly
(\ref{anomaly})). For a conformally flat metric with the torus
topology they are both vanishing. Therefore, our regularized
one-loop action actually does not have a typical renormalization
ambiguity quadratic in the curvature -- the term of the same
structure as $E_0\,\eta_0$ in (\ref{1000}), $\int d^4x g^{1/2}
R^2\sim\int d\tau a^3/a^4$. Thus, the vacuum energy in an Einstein
static spacetime (\ref{E_0}) should be uniquely calculable. This was
independently confirmed by different methods \cite{E_0}, and it
equals
    \begin{eqnarray}
    m_P^2\,E_0=\frac1{960}\times\left\{\begin{array}{c} 4 \\
    17\\
    88\end{array}\right.     \label{Casimir}
    \end{eqnarray}
respectively for scalar, spinor and vector
fields.\footnote{\label{footnote}This is the second major point of
departure from \cite{Tye1,Tye23}. Covariant regularization of the
vacuum energy never generates terms of the form $(M_s^4/H^4)\int
d\tau/a$ in which the structure $\int d\tau/a$ is accompanied by
powers of ultraviolet $M_s^4$ and infrared $H^2$ cutoffs advocated
in \cite{Tye1,Tye23}. This structure arises for generic
non-conformal fields from logarithmically divergent
curvature-squared terms only in the combination $\ln(M_s^2/H^2)\int
d\tau/a$, because powers of the ultraviolet cutoff can only
renormalize effective $\Lambda$ ($\int d\tau\,a^3$ structure) and
$G$ ($\int d\tau\,a$ structure) and do not accompany radiation-type
terms. But logarithms are too weak in the infrared to compete with
the tree-level exponential of (\ref{1}).} It should be emphasized
that (\ref{Casimir}) is the renormalized vacuum energy obtained from
(\ref{E_0}) by subtracting the quartic and quadratic divergences.
The result of this renormalization is a {\em positive} Casimir
energy (which is positive also for the naively negative vacuum
energy of a spinor field $-\sum_n(\omega_n/2)$, \cite{E_0}).

Finally, we have to take into account the effect of the finite
ghost-avoidance renormalization (\ref{renormalization}) which should
be applied also to $\Gamma_{\rm 1-loop}[\,\bar g\,]$. Since $\int
d^4x\,\bar g^{1/2}\bar R^2=72\,\pi^2 \eta_0$, this renormalization
leads to the expression similar to (\ref{1000}), but with the
modified vacuum energy which we denote by $C_0$
    \begin{eqnarray}
    &&\Gamma_{R}[\,\bar g\,]
    =m_P^2\,C_0\,\eta_0+F(\eta_0),          \label{GammaRR}\\
    &&m_P^2\,C_0=m_P^2\,E_0
    +\frac3{16}\,\alpha.                       \label{GammaR}
    \end{eqnarray}

One can directly check by using (\ref{100}) and (\ref{Casimir}) that
for all conformal fields of low spins the modified energy reduces to
the {\em one half of the coefficient} $B$ in the conformal part of
the total effective action (\ref{BmP^2})\footnote{This result
implies that the vacuum energy of conformal fields in a static
Einstein universe can be universally expressed in terms of the
coefficients of the conformal anomaly
$m_P^2\,E_0=3(2\beta-\alpha)/16$.}
    \begin{eqnarray}
    m_P^2\,C_0=\frac12\,m_P^2\,B.         \label{universality}
    \end{eqnarray}
This relation is confirmed by $\zeta$-functional and
point-separation regularizations, but not the dimensional one. The
latter results in the alternative value of $\alpha$ for a vector
field, $\alpha=-2/15$ instead of $1/5$ in (\ref{100}), and also
gives different results for the renormalized vacuum energy. This
lack of universality casts certain doubt on the dimensional
regularization, and in what follows we will accept the values
consistently obtained within other regularization schemes. They
maintain the relation (\ref{universality}) which will be very
important below.

\section{Effective Friedmann and bootstrap equations}
Now we assemble together the classical part of the action,
contributions of the conformal anomaly (\ref{correction}) and of the
static instanton (\ref{GammaRR}). By rewriting the conformal time as
a parametrization invariant integral in terms of the lapse $N$ and
the scale factor $a$,
    \begin{equation}
    \eta_0 = 2\int_{\tau_-}^{\tau_+}
    \frac{d\tau\,N(\tau)}{a(\tau)},                \label{fulltime}
    \end{equation}
we finally have the full effective action
    \begin{eqnarray}
    &&\Gamma[\,a(\tau),N(\tau)\,]=
    2 m_P^2\int_{\tau_-}^{\tau_+} d\tau\left(-\frac{a\dot{a}^2}N
    - Na + N H^2 a^3\right)\nonumber\\
    &&\qquad\qquad\qquad\qquad
    +2B m_P^2\int_{\tau_-}^{\tau_+}
    d\tau \left(\frac{\dot{a}^2}{Na}
    -\frac16\,\frac{\dot{a}^4}{N^3 a}\right)\nonumber\\
    &&\qquad\qquad\qquad\qquad
    +F\left(2\int_{\tau_-}^{\tau_+}
    \frac{d\tau\,N}{a}\right)+ B m_P^2
    \int_{\tau_-}^{\tau_+}
    \frac{d\tau\,N}{a}\,.                      \label{Gamma}
    \end{eqnarray}
It is essentially nonlocal due to the nonlinear dependence of the
free energy term on the conformal time (\ref{fulltime}), which
incorporates the "breathing" of the instanton in $\tau$-direction
caused by local variations in $a(\tau)$ and $N(\tau)$.

The effective Friedmann equation obtained from this action by
varying the lapse reads as
    \begin{eqnarray}
    &&\frac{\delta\Gamma}{\delta N}=
    2m_P^2\left(\frac{a\dot{a}^2}{N^2}
    - a + H^2 a^3\right)
    \nonumber\\
    &&\qquad\qquad\qquad
    +2Bm_P^2\left(-\frac{\dot{a}^2}{N^2 a}
    +\frac12\,\frac{\dot{a}^4}{N^4 a}\right)
    +\frac2a \left(\frac{dF(\eta_0)}{d\eta_0}+\frac{B}2
    m_P^2\right)=0.
    \end{eqnarray}
One can easily check that the variation of $a$ does not result in
the new equation because of the one-dimensional covariance of the
action amounting to the Noether identity
    \begin{eqnarray}
    \frac{\dot a(\tau)}{N(\tau)}\,
    \frac{\delta\Gamma}{\delta a(\tau)}=
    \frac{d}{d\tau}\frac{\delta\Gamma}{\delta N(\tau)}.
    \end{eqnarray}

In the gauge $N=1$ the effective equation reduces to the Friedmann
equation (\ref{Friedmann0}) modified by the quantum $B$-term
    \begin{eqnarray}
    \frac{\dot{a}^2}{a^2}
    +B\,\left(\frac12\,\frac{\dot{a}^4}{a^4}
    -\frac{\dot{a}^2}{a^4}\right) =
    \frac{1}{a^2} - H^2 -\frac{C}{ a^4},     \label{Friedmann}
    \end{eqnarray}
in which the radiation {\em constant} $C$ is a {\em functional} of
$a(\tau)$, determined by the {\em bootstrap} equation
    \begin{equation}
    m_P^2 C = m_P^2\frac{B}2 +\frac{dF(\eta_0)}{d\eta_0}. \label{bootstrap}
    \end{equation}
Here
    \begin{equation}
    \frac{dF(\eta_0)}{d\eta_0}
    =\sum_{n=1}^{\infty}\frac{n^3}{e^{n\eta_0}-1}  \label{thermalenergy}
    \end{equation}
is the thermal energy of a hot gas of particles, which adds to their
vacuum energy $m_P^2 B/2$. Thus, the overall back reaction effect is
mediated by the contribution of the radiation-type energy density
term and the anomalous quantum $B$-term. In view of Eqs.
(\ref{fulltime}) and (\ref{bootstrap}) the constant $C$
characterizing the amount of radiation nonlocally depends on the FRW
background supported by the radiation itself, and this is the
mechanism of the bootstrap we are going to analyze.\footnote{Note
that, contrary to a conventional wisdom, the vacuum energy of
quantum fields in a static universe (\ref{Casimir}) (and its
renormalized version (\ref{GammaR})) contributes {\em not} to the
cosmological constant, but to the total radiation density. This
follows from the structure of the corresponding terms in the
effective action discussed in Sect.5, cf. footnote \ref{footnote},
and is also confirmed by the radiation-type equation of state
derived for $E_0$ in \cite{E_0}.}

The on-shell action $\Gamma_0$ on solutions of this bootstrap
system can be cast into a convenient form by expressing the
combination $-a+H^2 a^3$ in (\ref{Gamma}) in terms of other pieces
of the effective Friedmann equation (\ref{Friedmann}). Then, after
the conversion of the integral over $\tau$ into the integral over
$a$ between the turning points $a_\pm$ (if any) the on-shell action
takes the form
    \begin{eqnarray}
    \Gamma_0= F(\eta_0)-\eta_0\frac{dF(\eta_0)}{d\eta_0}
    +4m_P^2\int_{a_-}^{a_+}
    \frac{da \dot{a}}{a}\left(B-a^2
    -\frac{B\dot{a}^2}{3}\right).              \label{action-instanton}
    \end{eqnarray}
The structure of the integral term here will be of crucial
importance for the elimination of the infrared catastrophe and
formation of the bounded cosmological landscape.

\section{The effect of conformal anomaly and bootstrap}
The modified Friedmann equation (\ref{Friedmann}) can be solved for
$\dot a^2$ as
    \begin{equation}
    \dot{a}^2 = \sqrt{\frac{(a^2-B)^2}{B^2}
    +\frac{2H^2}{B}\,(a_+^2-a^2)(a^2-a_-^2)}
    -\frac{(a^2-B)}{B}.          \label{time-der1}
    \end{equation}
Therefore, it has the same two turning points $a_\pm$ as in the
classical case (\ref{apm}) provided\footnote{Only the positive sign
of the square root is admissible in (\ref{time-der1}) because
otherwise the Euclidean configuration exists only outside of the
domain $a_-<a<a_+$, which violates the periodicity of the instanton
solution.}
    \begin{equation}
    a_-^2 \geq B,                   \label{require}
    \end{equation}
because in the opposite case the second term of (\ref{time-der1}) at
$a_-$ is positive and $a(\tau)$ at the contraction phase without
obstacle crosses the value $a_-$. The scale factor continues
decreasing and either becomes complex, at
$a=[B(2C-B)/(1-2BH^2)]^{1/4}$, or reaches zero. In the first case
the instanton solution is not periodic and should be excluded,
whereas the second case exactly corresponds to the formation of the
pure-state instanton of the Hartle-Hawking type. Indeed, the full
conformal time tends to infinity, $\eta_0\to\infty$, because of the
divergence of the integral (\ref{fulltime}) at $a(\tau)\to
a(\tau_-)=0$, so that both $F(\eta_0)\sim-\exp(-\eta_0)$ and
$dF(\eta_0)/d\eta_0$ vanish. Therefore, from the bootstrap equation
$C=B/2$, and the instanton smoothly closes at $a=0$, because in view
of (\ref{time-der1})
    \begin{equation}
    \dot{a}^2 \to 1+\sqrt{1-2C/B}=1,\,\,\,a\to 0. \label{velocity-lim}
    \end{equation}

If it were a conical singularity, the instanton solution could have
been ruled out for reasons similar to those discarding the
Hawking-Turok instantons \cite{HT}. For smooth instantons this
argument does not apply. Nevertheless, smooth Hartle-Hawking
instantons arising outside of the domain (\ref{require}) are ruled
out by the {\em infinitely large positive} value of their action.
For their solutions the action $\Gamma_0$ reduces to the last
integral term of (\ref{action-instanton}) with the lower limit
$a(\tau_-)=0$. Due to the contribution of the conformal anomaly and
in view of (\ref{velocity-lim}) its integrand is positive at $a\to
0$, and the integral diverges at the lower limit to $+\infty$. This
is the mechanism of how thermal fluctuations destroy the
Hartle-Hawking pure state and make a mixed state of the Universe
dynamically more preferable.

\begin{figure}[h]
\centerline{\epsfxsize 15cm \epsfbox{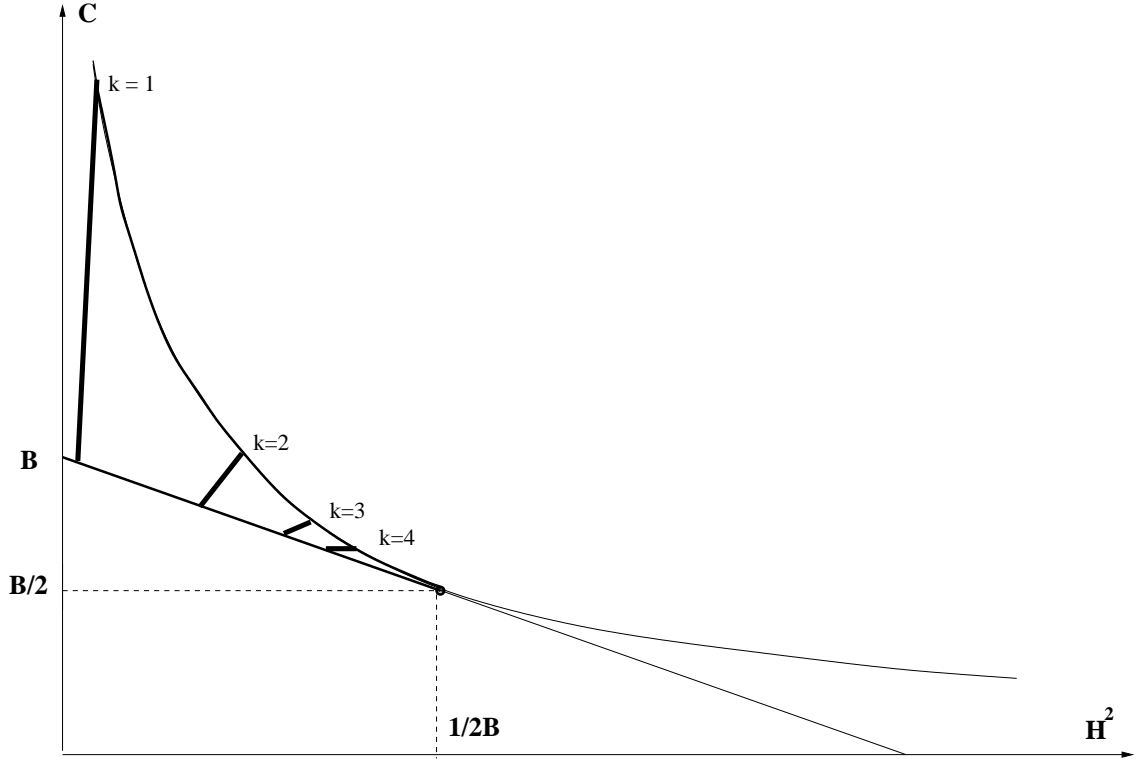}} \caption{\small The
instanton domain in the $(H^2,C)$-plane is located between bold
segments of the upper hyperbolic boundary and lower straight line
boundary. The first one-parameter family of instantons is labeled by
$k=1$. Families of garlands are qualitatively shown for $k=2,3,4$.
$(1/2B,B/2)$ is the critical point of accumulation of the infinite
sequence of garland families. \label{Fig.4}}
\end{figure}

Together with the upper bound on $CH^2$ (\ref{inequality}), the
condition (\ref{require}) is equivalent to the following
inequalities
    \begin{equation}
    4CH^2\leq 1,\,\,\,\,C \geq B-B^2 H^2,
    \,\,\,\,B H^2\leq\frac12,                     \label{restriction1}
    \end{equation}
which automatically imply that $C\geq B/2$. Thus, the admissible
domain for instantons reduces to the curvilinear wedge of the
$(H^2,C)$-plane below the upper hyperbolic boundary and above the
lower straight line boundary on Fig.\ref{Fig.4}. This domain lies to
the left of the critical point $C=B/2$, $H^2=1/2B$, at which these
two boundaries touch and form a cusp.

Inside this domain our bootstrap eliminates the infrared catastrophe
of $H^2\to 0$.
Indeed, the conformal time (\ref{fulltime}) can be rewritten as an
integral over $a$ between the turning points $a_\pm$. In terms of
the rescaled integration variable $x=Ha$ it reads as
    \begin{eqnarray}
    &&\eta_0 =\eta_0(q,p)=
    2\int_{x_-}^{x_+}
    \frac{dx}{x\Big(\sqrt{\frac{(x^2-p)^2}
    {p^2}+\frac{2}{p}\,
    (x_+^2-x^2)(x^2-x_-^2)}
    -\frac{x^2-p}{p}\Big)^{1/2}},  \label{conformal-mod1}\\
    &&x_\pm=\frac1{\sqrt 2}\,
    \sqrt{1\pm\sqrt{1-q}},
    \end{eqnarray}
and, in fact, turns out to be the function of two dimensionless
parameters $q$ and $p$,
    \begin{eqnarray}
    q\equiv 4CH^2,\,\,\,\,
    p \equiv BH^2.                         \label{p-define}
    \end{eqnarray}
For $H^2\to 0$ ($q\to 0$) it diverges to infinity at least
logarithmically at the lower integration limit\footnote{In terms of
the integral over $a$ this is the divergence at $a_+\sim
1/H\to\infty$, because of the singularity of the rescaling $x=Ha$ at
$H\to 0$. Thus, as it should be, this is the infrared limit of
infinitely large universes.} $x_-\to 0$, $\eta_0\to\infty$. Then it
follows from the bootstrap equation (\ref{bootstrap}) that $C\to
B/2$, which is impossible because in view of (\ref{restriction1})
$C\geq B$ at $H^2=0$. Thus the one-parameter family of instanton
solutions in the admissible wedge of the $(H^2,C)$-plane never hits
the $C$-axes and can only interpolate between the points on the
boundaries of the domain (\ref{restriction1}) with positive values
of $H^2$.

For a conformal scalar field the numerical analysis gives such a
family starting from the point at the lower line boundary with the
parameters
    \begin{eqnarray}
    H^2\approx2.997 \,m_P^2,\,\,
    C\approx0.004\, m_P^{-2},\,\,\,
    \Gamma_0\approx-0.1559,    \label{lower-point}
    \end{eqnarray}
and terminating at the point on the upper hyperbolic boundary
    \begin{eqnarray}
    H^2\approx12.968 \,m_P^2,\,\,
    C\approx0.0193 \,m_P^{-2},\,\,\,
    \Gamma_0\approx-0.0883.   \label{upper-point}
    \end{eqnarray}
The action $\Gamma_0$ changes in this family non-monotonically --
it has the maximum $\Gamma_0\simeq -0.063$ approximately at
$\eta_0=5.03$, so that maximally probable are the end-point
configurations, though the variation in their probability weights is
qualitatively negligible.

The instanton (\ref{upper-point}) describes the creation of a static
Einstein Universe of constant size, $a=a_\pm=1/(\sqrt2 H)$, filled
by a hot gas of a conformally invariant field in an equilibrium
state with the temperature
    \begin{eqnarray}
    T=\frac1{a\eta_0}=\frac{H}{\pi\sqrt{1-2BH^2}},   \label{T}
    \end{eqnarray}
which is analytically calculable as a function of $H$, because in
the limiting case of $4CH^2=q\to 1$, $a_-\to a_+$, the conformal
time integral (\ref{conformal-mod1}) yields
    \begin{equation}
    \eta_0 = \sqrt{\,2\,(1-2H^2B)}
    \int_{a_-}^{a_+}\frac{da}
    {\sqrt{(a_+-a)(a-a_-)}}
    = \sqrt{2\,(1-2H^2B)}\,\pi.             \label{conf-lim1}
    \end{equation}

Remarkably, for a static Universe the Euclidean action $\Gamma_0$
reduces to the first two terms of (\ref{action-instanton}) which
form the Legendre transform of $F(\eta_0)$. The entropy of the full
system which includes coupled radiation and gravity, is given by the
Legendre transformation of $\Gamma_0$ with respect to the inverse
temperature $\beta=\tau_0\equiv 1/a\eta_0$. Therefore, it equals
$S=-F(\eta_0)$. Apparently, this result can be reconciled with the
known relation between the free energy $F$ (it differs by the factor
of $\beta$ from the conventional definition), energy $E$ and
entropy, $F=\beta E-S$, by noting that the total energy of a
spatially closed system is zero, $E=0$.

The above results for a scalar field with $m_P^2\,B_{\rm
scalar}=1/240$ can also be obtained for other spins. We present them
for a vector field with a much bigger value of the constant
$m_P^2\,B_{\rm vector}=31/120$. The lower and upper bounds on the
instanton family then read respectively as
    \begin{eqnarray}
    &&H^2\approx1.063 \,m_P^2,\,\,
    C\approx0.187\, m_P^{-2},\,\,\,
    \Gamma_0\approx-0.2797,    \label{lower-point-vector}\\
    &&H^2\approx1.229 \,m_P^2,\,\,
    C\approx0.203 \,m_P^{-2},\,\,\,
    \Gamma_0\approx-0.2334.   \label{upper-point-vector}
    \end{eqnarray}
Together with (\ref{lower-point}) they give the lower bounds on the
instanton range for scalars and vectors (\ref{3})-(\ref{4}) cited in
Introduction. Their comparison shows the tendency of decreasing
Hubble scale and growing $-\Gamma_0$ with the growth of spin (or
the parameter $B$).

The corresponding behavior of our bootstrap at $B\gg1$ can be
studied analytically. For this purpose it is enough to stay at the
hyperbolic boundary, $H^2=1/4C$, where the value of the conformal
time is given by (\ref{conf-lim1}). In view of the bootstrap
equation (\ref{bootstrap}) it equals
    \begin{eqnarray}
    \eta_0=2\pi\sqrt{\frac{F'(\eta_0)}{m_P^2B+2F'(\eta_0)}},
    \,\,\,\,
    F'(\eta_0)\equiv\frac{dF(\eta_0)}{d\eta_0}.   \label{1001}
    \end{eqnarray}
For growing $B$ the solution of this equation tends to zero,
$\eta_0\to 0$, so that one can approximate the high-temperature
statistical sums (\ref{freeenergy}) and (\ref{thermalenergy}) by
$F(\eta_0)\simeq-\pi^4/45\,\eta_0^3$ and
$F'(\eta_0)\simeq\pi^4/15\,\eta_0^4$. Therefore, from (\ref{1001})
one has the asymptotic behavior of the conformal time
$\eta_0\simeq\pi(4/15\,m_P^2B)^{1/6}$ really decreasing with the
growth of $B$. For static instantons at the upper boundary the
integral term of (\ref{action-instanton}) identically vanishes, and
the on-shell action reduces to the contribution of the thermal terms
    \begin{eqnarray}
    \Gamma_0\simeq -\frac{2\pi}{3\sqrt{15}}
    \,m_P\sqrt{B},\,\,\,\,\,
    m_P^2 B\gg 1.                     \label{15000}
    \end{eqnarray}
It is important that $\Gamma_0$ stays negative (even though no
infrared catastrophe exists any more) while its absolute value grows
with $B$. This property implies that the growing spin of quantum
matter (growing $B=3\beta/4m_P^2$, cf. (\ref{100})) makes the
probability of the underlying instanton higher. Also it will be
important for suppressing the contribution of Lorentzian
(overbarrier) configurations in extended formulations of quantum
gravity, briefly discussed in Conclusions.

Qualitatively, this behavior stays true also for instantons below
the upper boundary. Indeed, with a growing $B$ the lower straight
line boundary of (\ref{restriction1}) becomes steeper, the critical
point $(1/2B,B/2)$ slides along the hyperbolic boundary in the
upward direction. The whole family of bootstrap solutions drops
deeper into the curvilinear wedge of Fig.\ref{Fig.4} approaching
closer to this point, $(C-B/2)/(B/2)\sim (30 B)^{-1/3}\to 0$, in the
vicinity of which members of this family do not differ much from one
another (cf.
Eqs.(\ref{lower-point-vector})-(\ref{upper-point-vector})).

The limit of large $B$ is very important. As will be discussed in
Conclusions, this limit arises when ascending the hierarchy of
particle phenomenology with a growing spin and number of matter
multiplets. This limit corresponds to strong quantum corrections
mediated by a large contribution of the conformal anomaly. These
corrections lead to a rather unexpected result. The higher the value
of $B$, the stronger it truncates the instanton domain
(\ref{restriction1}) from above, $H^2<H^2_{\rm max}=1/2B$. This
property will be discussed in Conclusions as a possible mechanism of
reducing the landscape of string vacua.

\section{Instanton garlands}
\begin{figure}[h]
\centerline{\epsfxsize 8cm \epsfbox{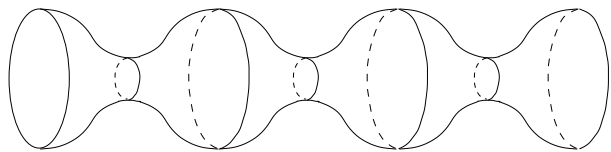}} \caption{\small
Segment of the garland consisting of three folds of a simple
instanton glued at surfaces of a maximal scale factor.
 \label{Fig.5}}
\end{figure}

The instanton family of Sect.7 does not exhaust the entire
cosmological landscape. Together with simple instantons of the above
type one can consider multiple folded instantons in which the scale
factor reaches its maximal and minimal values many times, similarly
to the instantons considered in \cite{mult-inst}. They can be
obtained by glueing together into a torus (at surfaces of the
maximal size $a_+$) $k$ copies of a simple instanton, see
Fig.\ref{Fig.5}. We will call them garlands. Their sequence will
generate a new quantum scale in gravity theory --- the upper bound
of the cosmological landscape (\ref{2}).
\\

The full lapse of the Euclidean conformal time for a $k$-length
garland equals
    \begin{equation}
    \eta_0^{(k)} =
    2k \int_{\tau_-}^{\tau_+} \frac{d\tau}{a}
    = 2k\int_{a_-}^{a_+}
    \frac{da}{a\dot{a}},           \label{conf-time-k}
    \end{equation}
and the whole bootstrap formalism persists with the replacement of
(\ref{fulltime}) by $\eta_0^{(k)}$ and multiplying the last
(integral) term of (\ref{action-instanton}) by $k$. The numerical
analysis for $k=2$ shows the existence of the instanton family
similar to the case of $k=1$. It interpolates between the point on
the lower boundary of $(C,H^2)$-plane
    \begin{eqnarray}
    H^2_{(2)}\approx45.89 \,m_P^2,\,\,
    C_{(2)}\approx0.0034\, m_P^{-2},\,\,\,
    \Gamma_0^{(2)}\approx-0.0113,    \label{lower-point2}
    \end{eqnarray}
and the point on the upper (hyperbolic) boundary
    \begin{eqnarray}
    H^2_{(2)}\approx61.12 \,m_P^2,\,\,
    C_{(2)}\approx0.0041\, m_P^{-2},\,\,\,
    \Gamma_0^{(2)}\approx-0.0145.    \label{upper-point2}
    \end{eqnarray}

Moreover, it turns out that such families exist for all $k$, $1\leq
k\leq\infty$, and their infinite sequence tends to the critical
point $C=B/2$, $H^2=1/2B$, where the lower and upper boundaries
merge. For each $k$ the instanton family forms a line joining the
upper and lower boundaries, the length of this line getting shorter
and shorter with $k\to\infty$ and its location closer and closer
approaching the critical cusp point. As in the case of $k=1$ the
upper point of each family gives rise to a hot static Universe
filled by radiation in the equilibrium state with the temperature
$T_{(k)}$ differing from (\ref{T}) by an extra $1/k$-factor.

The existence of this sequence of instanton families follows from
the behavior of the single fold of the conformal time
(\ref{conf-lim1}) at $2BH^2\to 1$. It tends to zero in this limit,
so that when multiplied by $k$ it admits the solution of the
bootstrap equation for any $k\to\infty$ with the total time
$\eta_0^{(k)}$ slowly growing to infinity. In this limit the
sequence of instanton families can be described analytically. For
simplicity consider the upper hyperbolic boundary where
$\varepsilon\equiv C-B/2$ tends to zero as we approach a critical
point and $H^2\simeq(1-2\varepsilon)/2B$. The full conformal time
(that is (\ref{conf-lim1}) times $k$) then behaves as
$\eta_0^{(k)}\simeq 2\pi k \sqrt{\varepsilon/B}$. In the assumption
that $\eta_0^{(k)}\to\infty$ for $k\to\infty$ the bootstrap equation
reads as
    \begin{eqnarray}
    m_P^2\varepsilon\simeq \exp(-\eta_0^{(k)}),
    \end{eqnarray}
because in this limit $F'(\eta_0)\simeq\exp(-\eta_0)$. It has the
solution, $\varepsilon\simeq B\ln^2 k^2/4\pi^2k^2$, which indeed
confirms that the conformal time is slowly growing to infinity,
$\eta_0^{(k)}\simeq\ln k^2\to\infty$. A similar analysis can be
performed for the lower boundary of the instantin domain. Thus, it
turns out that within the $1/k^2$-accuracy both upper and lower
points of each $k$-th family coincide and read as
    \begin{eqnarray}
    &&H^2_{(k)}\simeq \frac1{2B}
    \left(1 - \frac{\ln^2k^2}{2k^2\pi^2}
    \right),                            \label{h-tilde}\\
    &&C_{(k)}\simeq \frac{B}{2}
    \left(1+\frac{\ln^2k^2}{2k^2\pi^2}
    \right),                        \label{c-tilde}\\
    &&\Gamma_0^{(k)}\simeq
    -m_P^2B\,\frac{\ln^3 k^2}{4k^2\pi^2}.         \label{action-k}
    \end{eqnarray}
Thus, the lengths of instanton families (both in $H^2$ and $C$
directions) decrease as the second order of the $1/k^2$-expansion,
$1/k^4$, so that on Fig.\ref{Fig.4} they fit in ever narrowing wedge
near the critical cusp of $C=B/2=1/4H^2$. With a growing $k$ all
garlands become more and more static and cool down to zero
temperature
    \begin{eqnarray}
    T_{(k)}=\frac{1}{\sqrt B\,\ln k^2}\to 0.   \label{Tk}
    \end{eqnarray}

It is remarkable, that in contrast to the tree-level instantons of
\cite{mult-inst} the garland action is not additive in $k$, so that
with $k\to\infty$ it goes to zero rather than to $-\infty$. This
happens because a naively additive part of the action -- the
integral term of (\ref{action-instanton}) even when multiplied by
$k$ tends to zero with a growing number of folds. This in its turn
follows from the growingly static nature of garlands -- in terms of
the deviation from the critical point $\varepsilon$ we have
$a_+-a_-\sim\varepsilon$, $\dot a\sim\sqrt\varepsilon$,
$B-a^2\sim\varepsilon$, and the integral term behaves like
$k\,\varepsilon^{5/2}\sim(\ln k^2)^{5/2}/k^4$. Because of the
decreasing temperature (\ref{Tk}) the thermal part of
(\ref{action-instanton}) also decreases, but slower than the
integral term, and yields (\ref{action-k}).

Thus, infinitely long garlands do not dominate the instanton
distribution, but their existence is very important, because they
generate a new quantum scale --- the upper bound of the instanton
range (\ref{2}). Their sequence converges to the cold,
$T_{(\infty)}=0$, static instanton with the vanishing action
$\Gamma_0^{(\infty)}=0$, which realizes this scale as a maximal
possible value of the Hubble constant in the instanton landscape
    \begin{eqnarray}
    H^2_{\rm max}\equiv H^2_{(\infty)}=1/2B.   \label{Hmax}
    \end{eqnarray}
For a scalar and vector fields it gives respectively the values
$120\,m_P^2$ and $60/31\,m_P^2$ cited in Introduction,
(\ref{3})-(\ref{4}).

\section{Conclusions}
The picture of the cosmological landscape that we get from the back
reaction of quantum matter is very attractive. It consists of the
ensemble of universes in the limited range of the cosmological
constant, filled by a hot quasi-equilibrium gas of conformal field
particles. After nucleation from instantons these universes expand
in the Lorentzian regime. Eventually the radiation dilutes to a
negligible density when the cosmological constant $\Lambda=3H^2$
starts dominating and supports the inflationary stage, after which
we get a usual large-scale structure formation scenario. This
mechanism works also for static Einstein universes (originating from
the upper boundary of the instanton domain), because they also start
expanding as a result of their gravitational instability.

It is interesting that the Lorentzian version of the effective
Friedmann equation (\ref{Friedmann}) admits at late stages of
evolution, $a\gg 1/H$, two branches of the quasi-exponential
expansion with the Hubble parameters
    \begin{eqnarray}
    H_\pm^2=\frac1{B}\big(1\pm\sqrt{1-2BH^2}\,\big).
    \end{eqnarray}
The original Hubble constant $H$ should have a composite
(inflaton-induced) nature in order to decay at the exit from
inflation, so that $H_-^2\simeq H^2\to 0$, whereas the second branch
enters the acceleration stage with $H_+^2\to 2/B$. Unfortunately,
for hierarchy reasons this mechanism cannot play the role of dark
energy, because the needed value of $B$ should be $10^{120}\,
m_P^{-2}$. It is tempting to speculate, though, that at the
fundamental level the parameter $B$ is a moduli variable which is
sufficiently small in the early Universe, but becomes exponentially
big at late stages due to some dynamical compactification mechanism
of the braneworld type \cite{branes} (see also \cite{EQGbranes} for
branes in Euclidean quantum gravity). This would imply that the new
quantum gravitational scale, induced by garlands, is responsible for
the dark energy phenomenon. However, the detailed consideration of
this idea goes beyond the scope of this paper.

Our cosmological landscape belongs to the bounded range (\ref{2}).
Its infrared cutoff is provided by the radiation back reaction which
makes a mixed state of the early Universe dynamically more
preferable. Back reaction dynamically suppresses the pure state
Hartle-Hawking instanton by its infinitely large positive action
generated by the conformal anomaly. In essence, this implies a
solution to the long-standing problem of boundedness of the {\em
on-shell} cosmological action in Euclidean quantum
gravity and eliminates the infrared catastrophe of a small
cosmological constant (This should not be confused with the off-shell
indefiniteness of the Einstein action caused by the metric conformal
mode). This effect survives even in the classical
limit of the bootstrap equations
(\ref{Friedmann})-(\ref{bootstrap}), $B=0$, for which the lower
bound of (\ref{2}) can be obtained in an analytical form. In this
limit the conformal time (\ref{conformal-mod1}) for $q=4CH^2\to 0$
behaves as $\eta_0\simeq-\ln (q/64)-3 q\ln q/4\to\infty$. When
substituted into the bootstrap equation (\ref{bootstrap}) with
$F'(\eta_0)\simeq \exp(-\eta_0)$, it gives $H^2\simeq
16\,m_P^2\left[1+48\,m_P^2 C\ln(1/m_P^2C)\right]\to 16\,m_P^2$,
$C\to 0$. Therefore, the infrared catastrophe is ruled out already
at the classical level by the infrared cutoff which for a single
conformal scalar field equals $\Lambda_{\rm min}=48\,m_P^2$.

In contrast, the high-energy cutoff (\ref{Hmax}) is the quantum
effect of vacuum energy and conformal anomaly, which generates a new
scale in gravity theory. In view of the relation (\ref{BmP^2}) this
scale is determined by the inverse of the coefficient of the
topological Gauss-Bonnet term in the conformal anomaly
(\ref{anomaly}),
    \begin{eqnarray}
    \Lambda_{\rm max}=\frac{2}{\beta}\,m_P^2,\,\,\,\,\,
    m_P^2\equiv 3\pi/4G,              \label{top}
    \end{eqnarray}
and it tends to infinity in the classical limit $\beta\to 0$. In the
cosmological landscape this scale is realized as a limiting point of
the sequence of garland-type instantons.

Of course, we considered only the contribution of conformally
invariant fields of a few low spins. Other fields, gravitons in
particular, can change the picture, but it is suggestive that the
situation will qualitatively stay the same, because on static or
quasi-static backgrounds their behavior does not differ much from
their conformally invariant analogues. What is very critical for our
results is the value of the vacuum energy $E_0$ of quantum fields,
which after the ghost-avoidance renormalization yields a particular
contribution $C_0=B/2$ to the total constant $C$ in the bootstrap
equation (\ref{bootstrap}). Quite interestingly, this value
guarantees that the bootstrap solutions have a lower bound for $C$
exactly coinciding with the critical point $C=B/2=1/4H^2$ at which
the infinite sequence of garlands is accumulating. This peculiar
fact apparently implies a sound link between the renormalization
theory on curved spacetime background, ghost avoidance criterion and
quantum gravitational tunneling.

Conformally non-invariant fields are likely to destroy this
relation, and one can imagine three possibilities. If overall $C_0$
is less than overall $B/2$ the picture qualitatively stays the same.
In particular, an infinite sequence of garlands survives, though it
tends the critical point at a finite value of the conformal time
$\eta_0$ and, therefore, with a finite temperature. If $B>C_0>B/2$,
this infinite sequence becomes truncated at some maximal number of
instanton folds. Finally, if $C_0>B$ the infrared catastrophe of
$\Lambda\to 0$ occurs again -- the one-parameter family of
instantons necessarily hits the $C$-axes of the $(C,H^2)$-plane at
$C_0$. This point exactly solves the bootstrap equation and
infinitely dominates the rest of the partition function (at $H^2\to
0$ the upper integration limit in (\ref{action-instanton})
$a_+\to\infty$, and the dominant second term of the integrand at
this limit makes $\Gamma_0\to -\infty$). Which of these three
possibilities is realized is an open question deserving further
study.

Another open question concerns the normalizability of our partition
on the infinite set of instantons. One might think that it is not
normalizable because of infinite summation over garland folds $k$.
However, at least naively the total continuous measure of the
instanton range (\ref{2}) is finite because
$\sum_k\big(\Lambda^{(k)}_{\rm max} -\Lambda^{(k)}_{\rm
min}\big)\sim\sum_k(1/k^4)<\infty$. In order to have a definitive
conclusion, though, one must take into account a zero-mode
contribution to the actual measure and also estimate preexponential
factors. The latter reduce to the quantum mechanical functional
determinants of the Hessian of the effective action
$\Gamma[\,a(\tau),N(\tau)\,]$, which is a nonlocal operator
rather than a differential one. Their calculation is doable and will
be reported elsewhere \cite{progress}.

The boundaries of the instanton range (\ref{2}) depend on particle
phenomenology. For a single conformal or vector field they have
Planckian values (\ref{3})-(\ref{4}). Therefore, with the
nonperturbative back reaction taken into account, we still have the
issue of semiclassical validity of our model.  This problem can be
handled by adding new multiplets of quantum fields and raising their
spin. There is a simple scaling behavior in the number of fields
$N$. All terms in the bootstrap equation (\ref{bootstrap}) rescale
linearly in $N$,
    \begin{eqnarray}
    C\to NC,\,\,\,B\to NB,\,\,\,\,
    F(\eta_0)\to NF(\eta_0),                  \label{scaling}
    \end{eqnarray}
whereas $\eta_0$ given by (\ref{conformal-mod1}) depends only on
combinations $BH^2$ and $CH^2$. Therefore, the bootstrap remains
consistent when in addition to (\ref{scaling}) the value of $H^2$
scales down as $H^2\to H^2/N$. This justifies the semiclassical
approximation for large $N$ with the relevant $1/N$-expansion.

Moreover, as was noted in the end of Sects.7 - 8, the growing spin
of a conformal particle (or the parameter $B=3\beta/4\,m_P^2$) also
lowers the landscape scale and in view of (\ref{15000}) makes its
probability weight higher. This means that when ascending the
hierarchy of spins, this scaling  in $N$ and $B$ reduces the domain
(\ref{2}) to more and more narrow subplanckian range and suggests a
long-sought selection mechanism for the landscape of string vacua.
Modulo the details of a relevant $4D$-compactification, this
mechanism might work as follows. For $\beta$ rapidly growing with
$N$, $\beta\sim N$, and with the spin, cf. Eq. (\ref{100}), the
upper scale (\ref{top}) decreases towards the increasing
phenomenology scale, and converges to the latter at the string scale
$m_{\rm s}^2$, $1/B\to m_s^2$, where a positive $\Lambda$ might be
generated by the KKLT or KKLMMT-type mechanism \cite{KKLT}. Our
conjecture is that at this scale our landscape of hot instantons
selects from the enormous landscape of string vacua a reasonably
small subset compatible with the observed particle phenomenology and
large-scale structure. The corresponding bootstrap is perturbatively
consistent provided $1/B=m_{\rm s}^2\ll m_P^2$.

Our results hold within Euclidean quantum gravity incarnated in the
path integral (\ref{rho}) over Euclidean configurations. It gives
rise to the Lorentzian spacetime by analytical continuation but,
otherwise, automatically excludes Lorentzian saddle point
configurations. However, one can imagine an extended formulation of
quantum gravity generalizing the path integral (\ref{rho}) to a
wider integration domain or using an alternative formalism, like the
Wheeler-DeWitt equation. Within such an extension Lorentzian
solutions can, in principle, exist above the upper hyperbolic
boundary of (\ref{restriction1}), in the domain $4CH^2>1$
overbarrier from the viewpoint of the minisuperspace Wheeler-DeWitt
equation. They will contribute to the cosmological density matrix,
and this casts certain doubt on the universality of our results. Our
conclusions, however, are likely to stay true. Indeed, according to
(\ref{15000}) for a high value of $B$ the effective action scales as
$\Gamma_0\!\sim\!-m_P\sqrt B$. Therefore, because it is still
{\em negative} our Euclidean landscape at the string scale mentioned
above is weighted by $\exp(\#m_P\sqrt B)\!=\!\exp(\#\,m_P/m_{\rm
s})\gg 1$. Therefore, it exponentially dominates over Lorentzian
configurations, the amplitudes of which being $O(1)$ in view of
their pure phase nature. Thus, our results look robust against
possible generalizations of Euclidean quantum gravity.

This is how cosmological landscape emerges from nothing and perhaps
tames its string counterpart, provided some like it hot.

\ack{
A.B. thanks H.Tye, A.Gorsky, R.Woodard, S.Solo\-dukhin and,
especially, Jim Hartle for thought provoking discussions and helpful
suggestions. He is also grateful for hospitality of the Physics
Department of the University of Bologna. The authors are grateful
for hospitality of the Theoretical Physics Institute of the
University of Cologne where this work has been accomplished. The
work of A.O.B. was also supported by the Russian Foundation for
Basic Research under the grant No 05-02-17661 and the grant
LSS-4401.2006.2. A.K. was partially supported by RFBR grant
05-02-17450 and by the grant LSS-1757.2006.2.

\end{document}